\documentclass[letterpaper,twocolumn,10pt]{article}
\usepackage{usenix}

\usepackage{graphicx}
\usepackage{array}        %
\usepackage{booktabs}
\usepackage{amsmath}
\usepackage{amssymb}       %
\usepackage{subcaption}
\usepackage{xspace}
\usepackage{pifont}        %
\usepackage{balance}          %
\usepackage{xurl}             %

\graphicspath{{figures/}}

\newcommand{\sys}{HBFSim\xspace}

\newif\ifanon
\anonfalse

\newcommand{\hbfcallout}[4]{%
  \par\addvspace{0.6ex}%
  \refstepcounter{#1}%
  \noindent\textbf{#2~\csname the#1\endcsname:}\enspace\emph{#4}%
  \label{#3}%
  \par\addvspace{0.6ex}%
}
\newcommand{\observationbox}[2]{\hbfcallout{observation}{Observation}{#1}{#2}}
\newcommand{\challengebox}[2]{\hbfcallout{challenge}{Challenge}{#1}{#2}}

\newcommand{\chref}[1]{Challenge~\ref{#1}}

\begin{document}

\date{}

\title{\Large \bf HBFSim: Fast and Faithful Simulation of High-Bandwidth
Flash Under Real GPU Execution}

\ifanon
  \author{}
\else
  \author{
  {\rm Yanpeng Hu$^{1}$\thanks{Equal contribution.}\quad
   Yiwei Yang$^{2}$\footnotemark[1]\quad Yuanwu Zhu$^{3}$\quad
   Kexin Chu$^{4}$}\\[0.30em]
  {\rm Yusheng Zheng$^{2}$\quad Andi Quinn$^{2}$\quad
   Wei Zhang$^{4}$}\\[0.45em]
  $^{1}$ShanghaiTech University \quad $^{2}$UC Santa Cruz\\[0.20em]
  $^{3}$University of Science and Technology of China \quad
  $^{4}$University of Connecticut\\[0.35em]
  {\small\tt huyp@shanghaitech.edu.cn \quad
   \{yyang363,\,yzhen165,\,aquinn1\}@ucsc.edu}\\
  {\small\tt zzyuanyi@mail.ustc.edu.cn \quad
   \{kexin.chu,\,wei.13.zhang\}@uconn.edu}
  }
\fi

\maketitle

\begin{abstract}

High-Bandwidth Flash (\textbf{HBF}) places high-capacity NAND beside HBM to relieve the memory-capacity bottleneck of LLM inference, yet its system-level behavior cannot be evaluated before hardware becomes available.  Cycle-level GPU simulators are too slow for production-scale models.  Trace replay has a further shortcoming: it cannot capture the allocation, migration, and execution changes induced by different HBM--HBF configurations.  Our key insight is that HBF need not be evaluated by simulating the GPU: only the program-visible effects of HBF need to be modeled.  And only a real LLM workload running on real hardware can answer the arguments about HBF.  Hence the modeled service has to be injected into that running program, and the injection must not destroy the GPU concurrency that would hide the original I/O latency.

We present \sys, an open-source HBF simulator that executes LLM workloads on a real GPU while modeling HBF timing, thermal, and other behaviors online.  \sys rewrites the PTX of the workload's kernels and routes accesses inside a registered address range into the HBF simulator.  It supports asynchronous TMA transfers and capacities beyond physical GPU memory.  \sys leaves the model's run unaffected across ordinary-memory, TMA, and capacity-mode tests.  The delay it injects matches the delay requested to within \textbf{0.152}\%.  We also design a coupled thermal module that puts HBF, HBM, and the GPU in one advanced package, which is important for answering how severe the hot throttling problem becomes after HBF runs for a long time.  Experiments with Qwen3-30B show how package heating, HBM--HBF allocation, and shared MoE demand jointly constrain the design space of future HBF accelerators.
\ifanon\else
The source code of \sys is available at \url{https://github.com/SlugLab/hbfsim/}.
\fi

\end{abstract}

\section{Introduction}
\label{sec:intro}

Serving a large language model (\textbf{LLM}) is limited more and more by memory
bandwidth and memory capacity~\cite{yuan2024roofline,pope2022scaling}.
High-bandwidth memory (\textbf{HBM}), DRAM stacked inside the accelerator package, is
what the industry built to serve that load.  But HBM capacity does not keep up
with the parameter counts of the models it serves, and HBM is held back by its high manufacturing cost and by its scarce
supply~\cite{sandisk2025factsheet,micron2026form10q}.

High-Bandwidth Flash (\textbf{HBF}) is a
response to those limits. Sandisk and SK hynix introduced the HBF
specification through the Open Compute Project (\textbf{OCP}) in August
2026, with Google and Tenstorrent taking part in
validation~\cite{sandisk2026hbfspecrelease}; it is the first open standard that
defines stackable NAND flash as a memory tier inside the package of an xPU.
HBF is not an interface upgrade of a conventional NVMe
SSD, but a new near-memory architecture, in which NAND dies and the compute
chip, a GPU, xPU, or NPU, are integrated in one
package~\cite{ocp2026hbfspec}.  Sandisk's first generation of HBF products will put
HBF together with HBM and the GPU in one advanced package over
UCIe~\cite{sandisk2026hbfspecrelease,ocp2026hbfspec}, reaching at least \textbf{8$\times$} the
capacity of HBM at the same price with roughly the same read
bandwidth~\cite{sandisk2025factsheet,ocp2026hbfspec}.  What HBF makes new is the
medium: the stacked die holds NAND rather than DRAM of HBM.  
Because the medium is NAND, HBF makes a write far more
expensive than a read, and every write uses up part of the limited number of
times a flash cell can be rewritten.  Hence, the data suited to HBF is
data written once and read many times: model weights, and the part
of the key-value cache that several requests share, such as a system prompt
every request repeats.  HBF also inherits the latency of its medium: a NAND
read takes on the order of microseconds, roughly two orders of magnitude
longer than an HBM access~\cite{ju2026tilelens,kouchi2021flash}.  The execution
path of a transformer is predictable enough that an LLM can issue the reads it
will need ahead of the computation that consumes them and hide HBF latency
behind that computation~\cite{sheng2023flexgen,wang2026flashaccel}.  HBF is not a replacement for HBM; HBF is a
new tier below HBM.

\begin{table*}[!t]
\centering
\footnotesize
\caption{Comparison of four I/O paths that can serve accelerators in the
foreseeable future.}
\label{tab:tiers}
\begin{tabular*}{\textwidth}{@{\extracolsep{\fill}}lllll@{}}
\toprule
& \textbf{1st Generation HBF} & \textbf{HBM4} & \textbf{eSSD (Kioxia GP1)} &
\textbf{CXL-attached memory} \\
\midrule
Medium & 3D NAND & DRAM & XL-FLASH Gen2 & DRAM \\
Capacity & 512\,GB/stack & 64\,GB/stack~\cite{jedec2025hbm4} &
$\sim$2\,TB/drive & $\sim$0.1--2\,TB/device \\
Read bandwidth & 0.4--3.0\,TB/s & $\sim$2\,TB/s~\cite{jedec2025hbm4} &
$\sim$14--28\,GB/s & $\sim$30--60\,GB/s \\
Interface & UCIe & JEDEC 2048-bit bus/stack & PCIe 6.0 & CXL over PCIe \\
Read latency & $\sim$20\,$\mu$s~\cite{ju2026tilelens} & 10--100\,ns & $\sim$10--100\,$\mu$s &
$\sim$1.7\,$\mu$s from a GPU~\cite{sano2023cxlgpu} \\
Read granularity & 64\,B~\cite{ocp2026hbfspec} & 32\,B & 512\,B & 64\,B \\
Write endurance & limited P/E cycles & near-unlimited & 50 DWPD &
near-unlimited \\
Retention, unpowered & not guaranteed~\cite{ocp2026hbfspec} & none &
3 months at 40\,$^\circ$C~\cite{jedec2016jesd218b} & none \\
\bottomrule
\end{tabular*}
\end{table*}

Opinion on HBF splits into two opposing camps.  SK hynix, Sandisk, and
Joungho Kim of KAIST, widely known as the ``father of
HBM''~\cite{paek2026fatherhbm}, hold that HBF can take over a large part of
the capacity problem that HBM carries
today~\cite{ha2026h3,sandisk2026hbfspecrelease,jung2025hbfwinner,kwon2026hbf2038}.
Micron and others hold the opposite, that NAND's own limits will hold HBF back
and leave it worse than the media already in
use~\cite{petrucci2026hbfallyouneed}.  The first HBF devices for inference are
expected to reach customers as samples in 2027, and the choices a system
designer faces cannot wait for HBF silicon to arrive.  Real
parts that could settle the argument do not exist yet, so the questions left
open and the disagreement both call for a simulator that favors neither side. 

One simple idea for building an HBF simulator is to add an HBF module to a GPU
simulator such as
GPGPU-Sim~\cite{bakhoda2009gpgpusim,khairy2020accelsim,sun2019mgpusim}. 
Another idea is to add HBF latency into a replayed GPU trace.  Both have a fatal
flaw.  The first cannot run a real LLM to completion in a practical amount of
time~\cite{pan2026nextgen}, and only the result of a real workload settles the
core objections raised
about the latency of HBF. The second also suffers from excessive simulation
time. Furthermore, the same static trace cannot dynamically simulate different
HBF-to-HBM ratios and architectures, so the trace-simulation method cannot quickly explore the optimal
HBF configuration for different workloads. 

We accordingly choose to run a real LLM model on a real GPU.
Building this platform runs into many difficulties.  First, earlier simulators have no interception point targeting HBF, because HBF is the first standard to package NAND, GPU, and DRAM---three different media---inside the same accelerator, and they are linked die to die over UCIe, so a device effect can only be written into the instruction stream of the program under test.  Second, how long one access takes depends on
the accesses that came before, while the timing model has to be evaluated inside
the GPU code of the workload itself, and a GPU executes far out of order, so
tracking a Tensor Memory Accelerator (\textbf{TMA}) instruction accurately and
matching it to the real latency that belongs to it is very difficult.  Third, HBF is NAND flash, which is sensitive to
temperature, and it shares a package with a GPU that draws high power throughout
the run, so the rate HBF sustains is set by temperature. We must simulate the thermal problem.

We present \textbf{\sys}, which is the first open-source HBF simulator with the workload under evaluation
executing on a real GPU.  The GPU layer is not simulated
at all: the workload runs as usual on a real GPU, and rewritten PTX, the
intermediate code NVIDIA's compiler emits, routes accesses inside a registered
address range into \sys.  The workload on \sys is also real: Qwen3-30B runs under
injected HBF timing and emits token identifiers identical to the baseline run.

\section{Background}
\label{sec:bg}

HBF places NAND flash dies inside the accelerator package through the UCIe
interface.  The dies are stacked vertically: multiple layers of 3D NAND flash
are joined by through-silicon vias (TSV) and
micro-bumps~\cite{yin2026hbfserving}.  A silicon interposer or another advanced
packaging method places the stack and the compute chip in one package.
The 16-die NAND stack gives HBF $\sim$3\,TB/s of read
bandwidth~\cite{ocp2026hbfspec}. But NAND also gives HBF two main drawbacks.
\ding{192}~NAND wears out: a block of NAND tolerates
a limited number of program/erase cycles and must be erased before it is
rewritten~\cite{ocp2026hbfspec}. Hence HBF cannot be written as many times as HBM, because continuous
writing consumes its limited program/erase cycles. 
\ding{193}~NAND is also slow next to DRAM: an HBF page read is estimated at
$\sim$20\,$\mu$s~\cite{ju2026tilelens}, against about 200\,ns for one
global-memory access on a GPU with HBM~\cite{luo2024hopper}. Either property would keep NAND out of a memory tier
if the workload wrote often and needed each byte on demand. Fortunately, the workload characteristics of an LLM hide both drawbacks
well.  We make two observations:

\observationbox{obs:readonly}{Almost every byte an inference server moves
through accelerator memory is read rather than written.}

Decode is an important LLM inference stage bounded by memory bandwidth rather
than arithmetic~\cite{yuan2024roofline}. 
A large part of the bandwidth
exchanged between HBM and the GPU is model weights. Once training is complete,
model weights are written far fewer times than they are
read~\cite{li2026hbfsucks,petrucci2026hbfallyouneed,wang2026flashaccel}. A
prefix of the key-value cache shared across requests, such as a system prompt
every request repeats, likewise has far more reads than
writes~\cite{juravsky2024hydragen}. System prompts also account for a large
share of the traffic in long-horizon agent runs. A talk at Hot Chips 2026
reports that about 93 percent of the bytes of a trillion-parameter model are
read-only~\cite{agrawal2026hbfaicompute,kennedy2026oxmiqhbf}. 
Our own calculation
is less optimistic, but read-only bandwidth still accounts for 40 to 50
percent of all traffic for Kimi K3 with a 1M-token context.  The derivation is
not expanded here, for reasons of paper length and because this work focuses on
the HBF simulator.
In short, much of the bandwidth an existing LLM workload consumes does not need
the repeated overwriting that HBM supports, and industry is paying for that.
Cheap NAND media happen to suit data written once and read many times.

\observationbox{obs:prefetch}{The predictable execution path of transformer
inference leaves a long lead time for prefetching.}

Transformer decoding walks the layers in a fixed order, so which weights a
token reads is settled before that token starts to compute. The architecture of
an LLM leaves a large time gap for prefetching. By our calculation on an H100
SXM5 from NVIDIA~\cite{grattafiori2024llama3,nvidia2024h100}, computing layer
$i$ leaves a lead time of about 130.2\,$\mu$s for reading the weights of
layer $i{+}1$, which is 6.5 times the assumed $\sim$20\,$\mu$s HBF page
read latency~\cite{ju2026tilelens}. 

Better still, we find that architectures
such as Google's TPU and Huawei's DaVinci
NPU~\cite{jouppi2017tpu,jouppi2021tpuv4i,liao2019davinci} leave a larger time
window for HBF to prefetch weights than an NVIDIA GPU does. The reason is that
an accelerator which orders transfers in the compiler predicts the data to
prefetch better than a GPU that relies on run-time concurrency, and so issues a
fetch earlier~\cite{jouppi2021tpuv4i}.
In short, the other main theoretical
drawback of HBF, its very high latency next to HBM, can be hidden by
prefetching early, so that the compute interval covers the I/O latency.

We list several I/O paths that can serve
accelerators in the foreseeable future. Table~\ref{tab:tiers} shows that HBF, with its
lower price, large capacity, and bandwidth close to HBM, would in theory play
an important role in solving the memory wall problem of LLMs.
We therefore decided to design an open-source, fast, and accurate HBF simulator
to further explore how HBF behaves under various workloads.

\section{Motivation}
\label{sec:motv}

We first think of placing an HBF module inside a GPU simulator such as
GPGPU-Sim~\cite{bakhoda2009gpgpusim,khairy2020accelsim,sun2019mgpusim,pan2026nextgen}.
We find, however, that a simulator built that way cannot simulate HBF faithfully.
\begin{table}[t]
  \centering
  \footnotesize
  \setlength{\tabcolsep}{4pt}
  \begin{tabular}{@{}lr@{}}
    \toprule
    \multicolumn{2}{@{}c@{}}{Accel-Sim on lavaMD~\cite{huerta2025parallel}} \\
    \midrule
    Wall-clock time, one thread           & $>$5 days \\
    Slowdown against hardware, one thread & 10,748,031$\times$ \\
    Slowdown against hardware, 16 threads & 766,423$\times$ \\
    \specialrule{1pt}{0.7ex}{0.45ex}
    \multicolumn{2}{@{}c@{}}{Simulated warp instructions per second:} \\
    \midrule
    Accel-Sim, trace mode~\cite{khairy2020accelsim}     & 12,500 \\
    Accel-Sim, execution mode~\cite{khairy2020accelsim} & 6,000 \\
    GPGPU-Sim 3.x~\cite{bakhoda2009gpgpusim,khairy2020accelsim} & 3,000 \\
    Accel-Sim 2.0~\cite{pan2026nextgen}                 & 27,500 \\
    \specialrule{1pt}{0.7ex}{0.45ex}
    \multicolumn{2}{@{}c@{}}{Simulated instructions per second:} \\
    \midrule
    MGPUSim, four threads~\cite{sun2019mgpusim}   & $\approx$27,000 \\
    GPGPU-Sim, same machine~\cite{sun2019mgpusim} & $\approx$800 \\
    \bottomrule
  \end{tabular}
  \caption{Simulation time and instruction rate reported for cycle-level GPU simulators.}
  \label{tab:simspeed}
\end{table}
Table~\ref{tab:simspeed} shows that cycle-level GPU simulators are too slow to
finish one LLM inference run in acceptable time.  The disagreements over HBF can
be answered forcefully only on a real running LLM model.  Beyond simulation time, the
timing parameters of HBF have not been published: the specification defers
every quantity with the dimension of time to each vendor's product datasheet,
leaving no parameters to fill in.

We next think of driving an HBF simulator from a replayed GPU trace, and that
construction does not work either.  Trace mode is still slow:
Table~\ref{tab:simspeed} shows 12,500 warp instructions per second, which cannot
keep pace with the dynamic instruction count of one real LLM inference.  More
fundamentally, one trace corresponds to one HBM-to-HBF capacity split.  Changing
the split changes which tensors the inference framework keeps resident, how it
partitions the KV cache, and whether migration kernels fire at all; none of
those instructions appear in an already-recorded trace.  A single trace
therefore cannot explore different HBF-to-HBM ratios or architectures, and
cannot quickly locate the optimal HBF configuration across workloads.

We therefore run the workload on a real GPU, so that the computation that may hide
an HBF access comes from the hardware itself.  A real GPU, however, leaves an
experimenter no model to modify and no simulated clock to advance; the three
challenges stated below follow from that decision.

\challengebox{ch:intercept}{Traditional simulators have no interception point
applied to HBF.}

In the traditional simulator methodology, an interception point is a place
outside the program under test through which every memory access must pass and
where an experimenter can add waiting time.  HBF is the first medium that places
NAND, GPU, and DRAM in one heterogeneous package, and every access to HBF is
served inside that package.  Memory-tier emulators such as
Cylon~\cite{yoon2026cylon} and CXLMemSim~\cite{yang2026cxlmemsim} intercept in
the host-side CPU and kernel.  Block-device emulators such as
FEMU~\cite{li2018femu} and SwarmIO~\cite{kim2026swarmio} intercept in the
host-side submission queue and driver.  On a unified-memory APU, the
interception point is host-side software that fills the page
table~\cite{wahlgren2025mi300a}.  External memory is intercepted on the link
outside the package~\cite{sano2023cxlgpu}.  An access to HBF, issued by the
GPU's load/store units, passes through none of these places.  The one route that
remains is to instrument the code of the program under test~\cite{villa2019nvbit}.

\challengebox{ch:timing}{Attributing a modeled delay to the instruction that
should have paid it is difficult.}

To answer Challenge~\ref{ch:intercept} we intercept at the PTX level.  PTX,
Parallel Thread Execution, is NVIDIA's virtual instruction set: the compiler
emits it for a GPU kernel, and a backend turns it into the machine instructions
the GPU executes.  We insert the waiting time by rewriting memory instructions
in PTX.  However, PTX interception raises a doubt about where the delay is
injected: the PTX text says only which instructions the program contains.  How
the hardware resolves dependences after issue, and at which instruction it
stops, cannot be read from the PTX text.  When a warp issues a memory
instruction, the hardware sends the access out and marks the destination
register as not yet returned in the scoreboard, the hardware table that records
which register results have not come back.  The warp then issues the
instructions that follow immediately, without waiting for the value.  A later
ALU instruction that does not read the destination register executes as usual.
When the consumer instruction reaches issue, the hardware has to read the
destination register.  The hardware finds the mark still in the scoreboard and
stops at that instruction until the value returns.  The latency itself occurs
when the access is issued.  Whether the program pays time for that latency is
decided at the consumer, by whether the value has arrived.  A wait inserted into
the instruction stream becomes time the program spends waiting only when it sits
at the consumer.  A wait at the issue point makes the program wait for a stretch
of time the hardware would not have waited.  A rewrite sees only the
instruction-issue front end, while a memory instruction acts on back-end
execution that issue never sees.  PTX fixes which instructions the program
contains.  The backend assigns the physical registers and schedules the
instructions only after the rewrite.  At the PTX level the physical registers
have not been allocated and the instruction schedule has not been fixed.  A
rewrite therefore does not know what the chain of register dependences looks
like.  A rewrite also does not know which instructions lie between the load and
its consumer.  For example, a program contains \texttt{LDG R8}, then $\ldots$,
then \texttt{IMAD R9 R8 R5 R6}.  The ellipsis stands for the many instructions
between the load and the multiply-add.  The load is issued and the hardware
issues past it.  The multiply-add reads \texttt{R8}, so it stops at issue until
the value arrives.

The doubt has two parts.  The first is the injection point.  A rewrite that
turns \texttt{LDG R8} into a call that blocks stops the whole warp at that
instruction.  The warp sleeps through the modeled delay.  Real hardware does not
wait at that instruction.  The ALU instructions that follow and do not read
\texttt{R8} execute as usual.  Blocking at the issue point turns the whole
modeled delay into time the program spends waiting.  Real hardware turns into
waiting only the part of the delay that outlasts the instructions in between.
The doubt is exactly this: a rewrite holds the instruction sequence that the
issue end sees, while the place where access latency becomes lost time is the
hardware backend.  Those two places are not the same place.  The second is
information.  A rewrite cannot see the schedule that \texttt{ptxas} finally
emits.  The rewrite therefore does not know how many instructions lie between
\texttt{LDG R8} and \texttt{IMAD R9 R8 R5 R6}, nor how long those instructions
take to run.  How long the window between the load and its consumer is cannot be
read from the PTX text.

The instructions between the load and its consumer decide how much of the
latency is hidden.  If both the real latency and the modeled latency fit inside
that window, the two runs behave identically; if only the real latency fits, the
simulated run stalls where the real run does not; if neither fits, the two
stalls have different lengths.  The Tensor Memory Accelerator (\textbf{TMA})
makes the simulation harder.  TMA is a unit on NVIDIA GPUs that moves a whole
tile of a multidimensional array asynchronously.  In PTX, a TMA transfer is a
single instruction; how many requests that transfer becomes, how it moves
through L2 cache, and which CTAs receive its multicast are all decided by
hardware below PTX.  In other words, on an architecture as out-of-order and
asynchronous as a GPU, keeping a simulation both accurate and cheap to run is
very challenging.
\challengebox{ch:thermal}{The medium of HBF is a stack of temperature-sensitive
NAND flash that shares a package with the GPU, so the rate HBF sustains over a
long run is set by temperature rather than by the nominal peak.}

If HBF's heat is not modeled, the wrong result is that the nominal peak
bandwidth is read as a rate the device can hold.  The three nominal
bandwidths in the specification, 0.4 to 3.0\,TB/s, need not be steady values in
real use: the workload itself raises the temperature past the thresholds the
specification defines, which throttles the device.  HBF, HBM, and the GPU share
one package, so the junction temperature---the temperature of the point inside a
chip that actually produces the heat, which is higher than the average
temperature of the package---sits above the average.  The coupled heating of
these three media, and the ceiling that coupling puts on the HBF rate, therefore
have to be modeled.  Temperature is not the only property of the NAND medium
that determines what HBF delivers over a long run: retention loss, read disturb,
and wear accumulate as the device is used, and the refresh the device performs to
keep its data takes bandwidth the workload's reads also need.

To meet the three challenges above, we present the architecture of \sys.
Section~\ref{sec:design} answers \chref{ch:intercept} with D1 and D5,
answers \chref{ch:timing} with D2, and answers \chref{ch:thermal} with
D3 and D4.

\section{Design}
\label{sec:design}

\begin{figure*}[t]
  \centering
  \includegraphics[width=\textwidth]{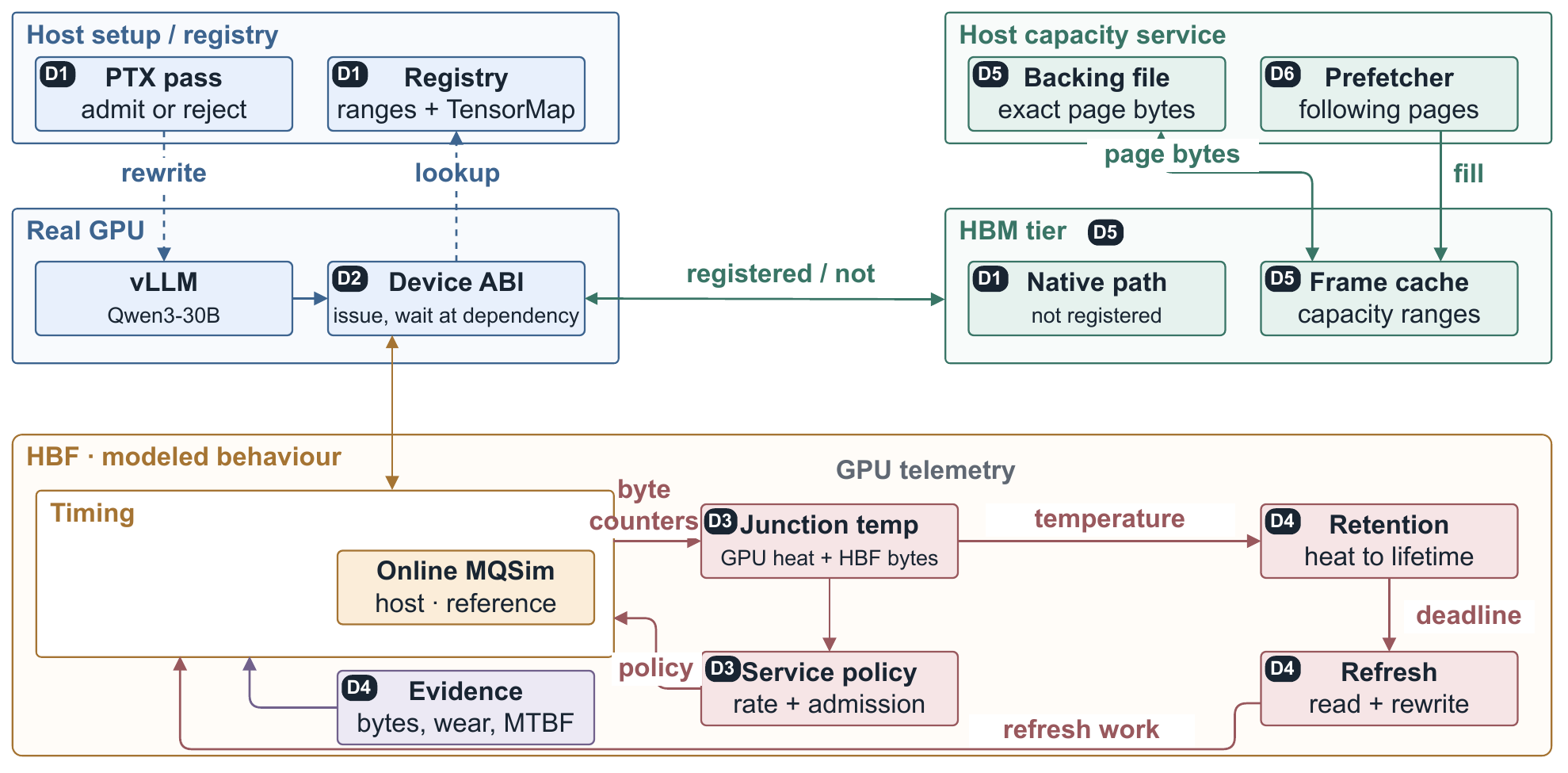}
  \caption{\sys architecture.}
  \label{fig:hbfsim-design}
\end{figure*}

\sys closes the loop between a running GPU program and a modeled HBF device.
The workload executes on an unmodified GPU software stack; a PTX pass rewrites
only accesses for which it can preserve the program's semantics.  At run time,
the generated helpers classify the effective address against explicit HBF
ranges and route the HBF portion through a detailed reference path, which places each foreground media request into an online MQSim instance and returns the completion to its original ring ticket.  That path consumes
the same generation-stamped thermal state and the same deterministic refresh
plan.  This section first
defines that execution boundary and then describes timing, temperature,
retention, and capacity as one coupled system.  Figure~\ref{fig:hbfsim-design}
shows where each part of that feedback loop executes.

\subsection{D1: An Explicit and Fail-Closed Boundary}

Applications register virtual-address intervals as HBF timing
or capacity ranges.  Registration is intentionally explicit: neither a PTX
instruction nor
a TensorMap descriptor says which physical tier will back its address at run
time.  Each range records its base, length, access permissions, page geometry,
stream, and backing-file offset.  The runtime publishes a sorted immutable
snapshot, and a device helper classifies the effective address produced by the
executing instruction.  An address outside every registered range follows the
native HBM path.  An in-range access is modeled only if the complete byte span
and its permissions are valid; an ambiguous or partially out-of-range access
fails rather than silently running as HBM.

This dynamic classification is essential for TMA
operations.  A TensorMap is a host-created descriptor whose base, dimensions,
strides, box shape, element format, and out-of-bounds policy are values, not PTX
constants.  The runtime binds a hash of the exact 128-byte descriptor to a
decoded and generation-stamped representation.  At issue, \sys enumerates the
tile's effective element addresses and partitions it into HBM, HBF, and
out-of-bounds bytes.  Unicast and multicast use the same partition; multicast
adds one tracked destination per selected cluster rank and completes its
barrier only after both native TMA and every modeled HBF segment complete.
Consequently, a tile that crosses the HBM--HBF boundary is modeled byte for
byte rather than assigned wholesale to either tier.

The same principle limits coverage.  The pass instruments PTX modules and
records an admission artifact containing the original and transformed hashes,
the helper ABI, and the instruction coverage.  A precompiled cubin, the already-built code for one GPU
architecture, has no PTX to rewrite, so \sys does not claim to observe it.  We
also measured the alternative: instrumenting the same run totally
through NVBit~\cite{villa2019nvbit}, which rewrites at the SASS level and reaches kernels
this pass cannot, costs more than ten times the runtime overhead of the PTX
pass, which makes simulating some larger-parameter LLMs on HBF take too long.
We therefore instrument at the PTX level.  Exact runs are admitted only
when every access required by the declared run contract is covered; otherwise
the launch is refused and the reason is recorded.  This turns incomplete
instrumentation from an implicit source of optimistic results into an explicit
experimental boundary.

\subsection{D2: Preserving GPU Asynchrony}

\textbf{Split issue from consumption.}  The naive idea is to make the rewritten
access wait in place: the program stops at the instruction that issues the
access and sleeps through the modeled latency.  Real hardware is the opposite:
it issues a load and continues to issue the instructions that follow, as long as
nothing later reads the register whose value has not returned.  If the rewritten
access waits in place, a single memory access stalls the whole warp at the issue
point, so the instructions that could otherwise overlap the HBF access never
issue.  We therefore split issue from consumption.  Issue is the moment the GPU
sends a memory access out, and consumption is the moment a later instruction
first reads the value the access returns.  Splitting issue from consumption
means handling the access at two separate places in the program.  At issue, a
rewritten access registers one record and continues.  The record holds the
modeled completion time of the rewritten access, and the design calls the
record a device future.  \textbf{At consumption, the thread waits for the
modeled completion time at the first instruction that actually reads the
value.}  Placing the wait at the point of consumption puts the wait in the same
place where the hardware itself stops.  The scoreboard is the hardware table
that tracks which register results have not yet come back.  When a memory access
is issued, the hardware marks the destination register as not yet returned in
the scoreboard.  The first instruction that reads that register finds the mark
still there and stops.

An ordinary load or store allocates a small device future and
continues; the first dependent use polls or waits for its modeled completion.
The workload does not run ahead of the model: a synchronous consumer
waits immediately, whereas a future or TMA object waits only at its
architectural dependency.
TMA issue creates a deferred object keyed by the barrier, CTA group, cluster,
and generation.  Native completion and modeled completion are conjunctive:
neither alone releases the barrier.  Every object has exactly one terminal
transition, and stale generations, leaks, and faults are reported.

\subsection{D3: A Closed-Loop Thermal Controller}

The CUDA process owns the live device identity and publishes GPU temperature,
power, a host-monotonic timestamp, and source status.  The source can instead
be an explicitly declared constant or hash-bound trace for deterministic
experiments.  Missing, stale, regressing, or failed telemetry terminates a
thermal run; \sys never freezes the last good temperature.  Host and GPU clock
origins are never subtracted.  Device code consumes only relative service and
refresh debt.

Media activity, not a nominal power figure, is the source of HBF heat in
\sys.  Each read, program, and erase event that the media model schedules
carries an energy under the event-based Joule energy model,
$E_{\mathrm{NAND}} = E_{\mathrm{command}} + e_{\mathrm{byte}} \times
\mathrm{bytes}$.  Event energies accumulate in 10\,ms bins.  The binning
applies a time-bin power integration: an event that spans more than one bin has
its energy split across the bins in proportion to the time overlap.  The power
of node $i$ in bin $k$ is $P_{i,k} = E_{i,k}/\Delta t + P_{\mathrm{idle},i}$.

\sys builds the package thermal operator offline with 3D-ICE
4.0~\cite{zhu2026threeice} from the material properties, the geometry, the
grid, and the power sources.  The temperature field then follows Fourier
transient heat conduction, $C_v\,\partial T/\partial t = \nabla\cdot(k\nabla T)
+ q'''$, where $q'''$ is the volumetric heat generation.  Discretized on the
grid, every thermal cell becomes an RC network node,

\begin{equation}
  C_i \frac{dT_i}{dt} = P_i - \sum_j G_{ij}(T_i - T_j)
  - G_{i,\mathrm{sink}}(T_i - T_{\mathrm{sink}}),
  \label{eq:thermal-rc}
\end{equation}

where $G_{ij}$ is the conductance between neighboring cells and
$G_{i,\mathrm{sink}}$ is the conductance from a cell to the heat sink.
Conduction between two cells is Fourier's law.  At the heat sink the operator
applies $-k\,\partial T/\partial n = h(T - T_{\mathrm{amb}})$, which is
Newton's law of cooling, or the Robin boundary condition in the finite-volume
formulation.  At a non-uniform bottom boundary the half-cell conduction and the
convection form a series thermal resistance with equivalent boundary
conductance $G_{\mathrm{eq}} = 2khA/(\Delta z\,h + 2k)$.

The GPU, the HBM, and the HBF share one conductance matrix $G$ and one
capacitance matrix $C$; heat from any of the three reaches the other two
through that operator, not through three separately solved models averaged
together.  At every 10\,ms step the runtime forms one package power vector from
the measured GPU power and the modeled HBM and HBF power, with separate entries
for the three media and the HBF layers.  The division of labor between \sys and
3D-ICE is as follows: 3D-ICE solves the underlying thermal physics; \sys owns
the energy balance of heat generation, heat transfer, heat storage, and heat
dissipation across the three media in one real run, and returns the resulting
package temperature to HBF request service.  Offline, each heat source receives
a 1\,W unit step, and the Eigensystem Realization Algorithm (ERA) fits a
reduced-order model to the recorded response.  At run time the reduced-order
model is a discrete state-space system,

\begin{equation}
\begin{aligned}
  x_{k+1} &= A x_k + B p_k + b,\\
  T_k &= C x_k + D p_k + d.
  \label{eq:thermal-era}
\end{aligned}
\end{equation}

The state $x_k$ is a reduced thermal state and not a physical grid temperature,
and the reduced model is far smaller than the full-order operator.

At steady state the reduced-order model gives the temperature change that one
source produces at every node.  Driving one source at a time with 1\,W fills
the thermal-resistance and cross-talk matrix $R_{ij} = \Delta T_i/P_j$, in K/W.
The HBF hotspot temperature is the output of the thermal model, and the
admission rate that the throttling state allows changes the media activity that
generated the heat.

It applies thermal service exactly once.  If an unthrottled operation has
service $s$ and the controller publishes service fraction $q$ in parts per
million, the device uses

\begin{equation}
  s_{T}=\left\lceil\frac{s\cdot 10^6}{q}\right\rceil .
  \label{eq:thermal-service}
\end{equation}

The fraction is snapshotted before queue reservation.  Thus a command admitted
before a state transition retains its service and drains once; a later Severe
state cannot retroactively strand it.  Normal publishes $q=10^6$, Light
publishes a profile value below $10^6$, Severe withholds new reservations, and
Shutdown returns a distinct terminal link-shutdown status.  A seqlock-style
generation protects the mode, service, temperature, and admission fields from
torn host--device reads.

The thresholds RTT, LTT and STT drive a hysteretic state machine with four states: Normal, Light, Severe and Shutdown.
The controller is constructed and publishes its first valid snapshot before the
daemon's readiness heartbeat, so no request can observe an uninitialized
thermal policy.

\subsection{D4: Retention, Refresh, and Wear as Injected Latency}

The difference from a conventional NAND simulator is not that \sys keeps these
states; it is where the number those states produce ends up.  A conventional
NAND simulator keeps them too: a refresh consumes time inside the device, and
the device latency the simulator reports grows.  That device, however, sits
behind a host interface, so however expensive a refresh is, it only moves a
quantity inside the device, and the program does not change what it does next
because of it.  Here the loop closes because the running program pays for the
refresh itself: a rise in temperature shortens the retention deadline, the
shortened deadline forces refresh writes, those writes take bandwidth the
workload's own reads also use and endurance the device has to spend, and the
next access the workload issues is injected with a longer delay.  In a
conventional simulator the chain temperature $\rightarrow$ retention deadline
$\rightarrow$ forced refresh $\rightarrow$ bandwidth and endurance
$\rightarrow$ a longer injected delay $\rightarrow$ a workload that feels it
breaks at the second-to-last step: it can compute the same refresh traffic, but
it cannot make the running program pay time for it.

Each registered block carries retention damage, a read-disturb
count, a zone
identifier, and current and maximum program/erase count (PEC).  Retention
lifetime follows an Arrhenius model anchored at the declared reference
condition:

\begin{equation}
 L(T)=L_{\mathrm{ref}}\exp\!\left[
 \frac{E_{a,r}}{k_B}\left(\frac{1}{T}-\frac{1}{T_{\mathrm{ref}}}\right)
 \right].
\end{equation}

Over a changing trajectory the controller accumulates

\begin{equation}
 D\leftarrow D+a_t\frac{\Delta t}{L(T)},
 \label{eq:retention-damage}
\end{equation}

where $a_t$ is a disclosed reliability-time acceleration used only to make
validation runs practical.  A block becomes eligible when damage reaches its
lead threshold or its read-disturb limit.  Eligibility never changes
application bytes.  The scheduler emits page-aligned read--rewrite pairs in a
deterministic order: eligibility epoch, channel round robin, die round robin,
zone, block, and page.  It excludes refresh from a die while an application
read occupies that die.

In the reference path, refresh uses a disjoint request-ID namespace and enters
the same MQSim queues as foreground traffic.  Its completion updates only
background state; it cannot consume an application ticket.  A block commits
one PEC and clears its damage and disturb counters only after every read and
rewrite quantum completes successfully.  On failure, the counters remain and
the work is eligible to retry.  The host publishes the same plan as relative
byte debt.  An application admission atomically claims at
most one refresh quantum and reserves the corresponding thermally scaled
service before its own service.  Host decay accounts separately for debt that
would have drained in otherwise idle time.  This representation avoids putting
host timestamps into GPU queue tails while preserving refresh bytes, order,
PEC, and contention with foreground traffic.

\subsection{D5: Capacity}

Timing ranges preserve their original address.  Capacity ranges
instead use a
bounded HBM frame cache backed by a sparse host file.  A miss hands a page to
the host service, which reads or programs the media model and copies the frame;
dirty eviction writes the exact bytes back.  Refresh reads and rewrites through
the backing abstraction without modifying payload contents, so the acceptance
condition is an unchanged workload checksum, not merely a completed media
request.

\subsection{D6: Speculative Prefetcher}

A dense LLM reads its weights in the same order on every token, so an HBF tier
can fetch the pages a layer needs before that layer runs.  Mixture-of-Experts
routing removes that order: a token's experts are chosen only after the
preceding layer produces its output.  D6 therefore predicts from the accesses
already served: \sys fetches the pages that follow a served page fault.  \sys
exposes a prefetch interface so that a user can supply a policy of its own, and
it also ships one simple default policy, the one described here.

On a synthetic Mixture-of-Experts stream, 8 pages per expert, 256 frames and
1033 distinct pages, the share of accesses served without a media read rises
from 17.97\% to \textbf{89.16\%}, and the media reads an access still has to
wait for fall by \textbf{86.79\%}, from 840 to 111.  Total media reads rise
from 840 to 979, or 16.55\%, and that rise is not the cost of fetching early:
729 (83.99\%) of the prefetcher's 868 reads each replaced a read that would
have happened anyway, issued earlier rather than in addition.  Only the 139
reads that went unused add traffic---pages guessed at an expert boundary, or
evicted before their access arrived---and 139 is exactly the rise.

The speculative prefetcher builds no media plan and never enters the detailed
timing path, so it adds no modeled latency, queueing, or contention.  Entering
the timing path is future work, and this paper claims no run-time performance
benefit from D6.

\section{Implementation}
\label{sec:impl}

The implementation is 9,302 lines of C, C++ and CUDA under \texttt{src/} and
\texttt{include/}, with a further 3,250 lines added afterwards for the measured
calibration work. Three parts are reused: the flash media model of MQSim, the
CUDA call interception path of bpftime, and the inference frameworks
themselves, whose source is not modified. The parts we add are the range table
and its registration interface, the PTX rewriting pass, the shared control
region and the interface across it, the calibrated timing model, the
thermal and retention model, and the page cache and backing store of the
capacity mode.
Four dependencies are pinned by commit: bpftime, MQSim, llama.cpp, and vLLM.
Patches are kept as separate
files instead of being applied into the imported source trees, so those trees
stay as they were fetched; each build artifact carries a record of the versions
that went into it, and the launcher refuses to start a build that carries no
such record or one whose record does not match. The GPU is an NVIDIA RTX PRO
6000 Blackwell Server Edition with driver 595.84, compute capability 12.0 ---
the version number CUDA uses to identify a GPU architecture generation --- and
97,887\,MiB of memory; the CUDA toolkit is 13.0.88, and PTX is assembled with
the ptxas of CUDA 12.8. The flash and thermal reference device is a Dell DC
NVMe CD8P E3.S 1.92\,TB on PCIe 5.0 at 32\,GT/s x4, attached to NUMA node 1; it
supplies real flash and thermal behaviour and is not an HBF device. Section~\ref{sec:eval} describes the same machine.

Host and device share one control region. It is carried in an anonymous memory
file, a region of memory that has a file descriptor but no file on disk, which
is then sealed against any change of size, mapped, and registered so that the
device can reach it, which yields a device-side address for the same bytes. The
seal is what makes the region usable from a GPU thread: the device side reaches
the request ring and the completion slots at fixed offsets, and those offsets
hold only if the size cannot change while the run is in progress. The other
allocation route was not taken because it does not yield a file descriptor that
can be mapped again after a change of process. A timed access reserves a
request slot and, at the same index, the completion slot that will answer it.
The reservation sequence number is written into the request and is also the
ticket the caller keeps; a waiter consumes only the completion slot its own
ticket names, so a completion is never taken by another access.

The daemon publishes its first heartbeat, a counter it updates to show that it
is still running, only after the configuration, the timing engine and the
dispatcher have been constructed. That first heartbeat is the readiness
boundary: until it appears, no access is served. One-time initialisation is
allowed 10 seconds, and from then on the heartbeat is updated at least once
every 10 milliseconds. The device side does not subtract a host clock reading
from a GPU clock reading, because the two clocks are not guaranteed to share a
time origin. It records the last heartbeat value it observed together with the
local GPU time at which that value changed, and decides from those two whether
the daemon is still running; a run whose heartbeat stops ends rather than
continuing untimed.

Four kinds of time are counted separately: the delay the model asked for, the
wait an access actually observed, the time the host spent serving it, and the
part of wall-clock time that the modeled delay does not account for. They are
carried in four counters, \texttt{modeled\_ns}, \texttt{wall\_ns},
\texttt{service\_ns} and \texttt{overhead\_ns}. The reason is that a simulator
which applies its
effects inside a real execution can be functionally correct and still spend
more time in its own software path than the device latency it is modeling; a
slowdown reported without that split says nothing about the device.
Section~\ref{sec:eval} reports the split for every workload.

Every failure ends the run, and nothing is downgraded silently. A request ends
in one of eight states: \texttt{PENDING}, \texttt{READY}, \texttt{IO\_ERROR},
\texttt{COPY\_ERROR}, \texttt{CHECKSUM\_ERROR}, \texttt{TIMEOUT},
\texttt{UNSUPPORTED} and \texttt{DAEMON\_LOST}; the six error states among them
end the run. When the dispatcher or the timing engine fails, two steps happen
in a fixed order. A terminal error completion is first posted to the reserved
ticket of every request that has been admitted, is in flight, or is still
queued, and only then is the global failure word published. In that order, a
waiter that observes the failure word also finds a terminal state in the
completion slot its own ticket names, so no thread is left waiting on a
completion that will never arrive.
At clean exit or any terminal thermal state, \sys atomically replaces one
canonical JSON summary.  It contains the exact profile and SHA-256, telemetry
source identity, temperature samples, every state transition, state residency,
application and refresh bytes, debt disposition, damage, PEC, and terminal
status.  It also reports a sensitivity sweep rather than a single asserted
lifetime.  For failure activation energy $E_a$, interval hazard is

\begin{equation}
 H(E_a)=\sum_i\frac{\Delta t_i}{M_{\mathrm{ref}}}
 \exp\!\left[\frac{E_a}{k_B}
 \left(\frac{1}{T_{\mathrm{ref}}}-\frac{1}{T_i}\right)\right],
\end{equation}

with failure probability $1-e^{-H}$.  Retention activation energy and
whole-device failure activation energy are separate profile inputs.  The
result is therefore a reproducible, parameterized model outcome---not a claim
that \sys measured the lifetime of unavailable HBF hardware.

\section{Evaluation}
\label{sec:eval}
\begingroup
\typeout{EV_DIMENSIONS: column=\the\columnwidth; text=\the\textwidth; height=\the\textheight}
We evaluate \sys before HBF hardware becomes available. Our evaluation addresses three questions:
\begin{itemize}
  \item \textbf{EQ1: Can \sys execute complete real LLM workloads transparently, preserving program results and synchronization, while applying modeled HBF latency accurately to the accesses it declares?}
  \item \textbf{EQ2: How does package heating affect HBF service, and which component reaches its thermal limit first?}
  \item \textbf{EQ3: How do the HBM--HBF mix and MoE routing together set the capacity a serving workload has?}
\end{itemize}

\begin{figure*}[!t]
\centering
\includegraphics[width=\textwidth]{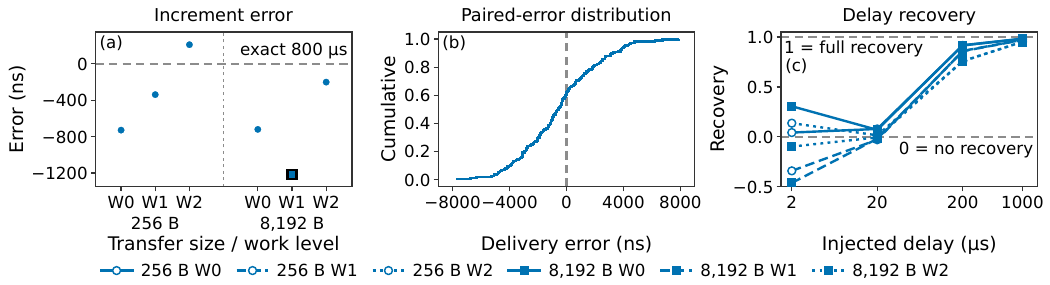}
\caption{\textbf{Timing-injection fidelity.} Increment error by configuration (a), paired-error distribution (b), and normalized delay recovery (c).}
\label{fig:eq1-timing-fidelity}
\end{figure*}

\begin{figure*}[!t]
\centering
\includegraphics[width=\textwidth]{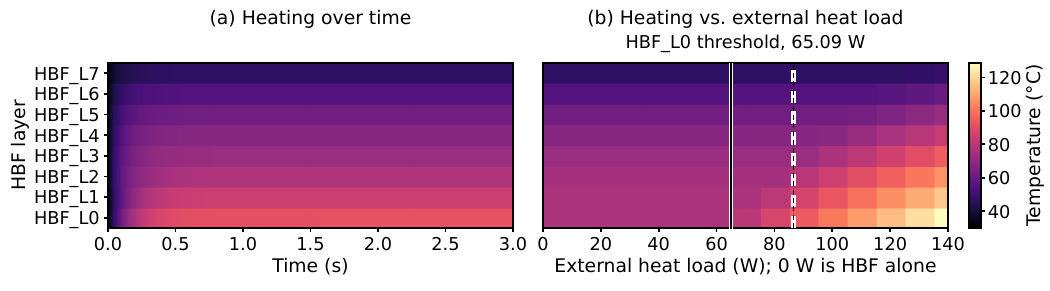}
\caption{\textbf{Modeled HBF layer temperatures.} Temperature of the hottest node in each of the eight HBF layers, over time (a) and across external heat loads (b), on one shared color scale.}
\label{fig:eq2-layer-heating}
\end{figure*}

\begin{figure}[t]
\centering
\includegraphics[width=\columnwidth]{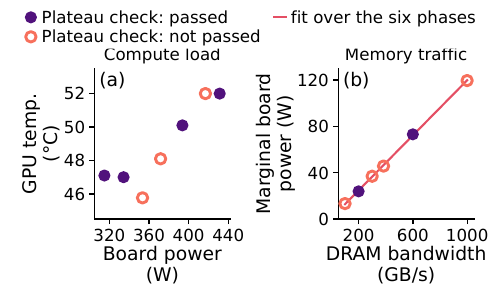}
\caption{\textbf{Physical reference on the evaluation GPU.} Board power against the GPU's own temperature sensor (a) and measured DRAM bandwidth against the board-power difference between a phase's DRAM-resident and L2-resident halves (b), over thirteen 300\,s phases.}
\label{fig:eq2-physical-reference}
\end{figure}

\begin{figure}[t]
\centering
\includegraphics[width=\columnwidth]{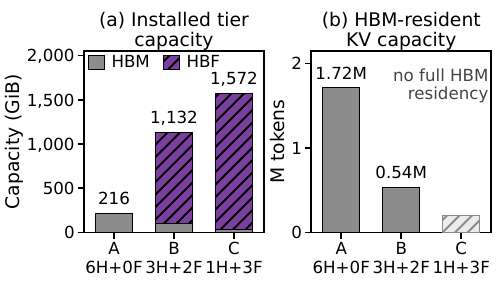}
\caption{\textbf{HBM--HBF capacity trade-offs.} Installed capacity (a) and HBM-resident key-value capacity (b) for three layouts under a common package-area budget.}
\label{fig:eq3-alloc}
\end{figure}
\subsection{Experimental Setup}
\label{subsec:eval-setup}\label{subsec:eval-current-setup}
GPU experiments run on the machine of Section~\ref{sec:impl}: an NVIDIA RTX PRO 6000 Blackwell Server Edition (compute capability 12.0, driver v595.84) with a Dell CD8P SSD. GDDR7 on the GPU is the platform's actual memory and backs the logical fast tier through a page cache. Qwen3-30B-A3B runs under vLLM v0.15.1, and the simulator is built with CUDA 13.0. The Dell CD8P is the reference for the measured flash and host-paging path only. The full-order 3D-ICE package model is the reference the service simulation's reduced model compared against, and the OCP specification supplies the HBF configuration constraints.

\subsection{EQ1: End-to-End Execution and Timing Fidelity}
\label{subsec:eval-eq1}\label{subsec:eval-fidelity}
\sys injects modeled HBF service into a real GPU run: the workload must compute what it would compute without instrumentation, and each access must be charged to the tier it actually reaches. An access charged to the wrong tier misattributes service, and breaking barrier handling changes GPU synchronization semantics. Completing a run and modeling a run are different claims: completing means the model loads, prefill and decode run, the requested output tokens are produced, and the process ends normally; modeling completely would mean every access the run declares should be modeled actually is. The question therefore asks four things of one run on Qwen3-30B-A3B: that it completes, that it computes what it would have computed without instrumentation, that the accesses reaching the modeled tier are the ones the registration declares and the rest keep native semantics, and that the service injected into them is timely.

Under vLLM v0.15.1, the instrumented build runs Qwen3-30B-A3B end to end at a fixed 32-token prompt, seed 0 and \texttt{output\_len} 8: a 60.988\,s load, then 44.469\,s of generation producing the eight requested output tokens; a TinyLlama run under llama.cpp takes ten 50\,ms injections.

The instrumented run preserves the second and third of those four things: value, effective address, tier assignment and synchronization are unchanged, over 42 checks that cover ordinary and TMA accesses across seven launchable TMA templates, in-range and out-of-range addresses, HBM--HBF mixed boundaries and barrier-dependent completion. A 110\,GiB logical range is mapped through a bounded 2\,GiB page cache, and the 128 checksums over the first and last regions touched equal the baseline's.

The timing result is a difference between two delays rather than an absolute match: a 1,000\,$\mu$s injection and a 200\,$\mu$s injection, an 800\,$\mu$s increment. Its six cells are TMA transfers at 256\,B and 8192\,B, and the records carry a constant 14.3--14.9\,$\mu$s offset in a segment the reference model does not contain. The two transfer sizes are crossed with three frozen work levels, each an iteration count of an integer work loop injected into the probe kernel alongside the outstanding transfer: 0, 3335 and 9769 at 256\,B, and 0, 3352 and 9787 at 8192\,B. The two non-zero levels were chosen so the in-kernel arithmetic lands about 10 and 30\,$\mu$s above the native transfer cost. Each cell is 36 paired blocks, and the reported cell measures 798,784.0\,ns at 8192\,B. The 0.152\% is the error of that increment in the reported cell, not an absolute HBF latency error: what the difference demonstrates is the injection and the measurement, and no physical HBF latency was measured (Figure~\ref{fig:eq1-timing-fidelity}a).

Figure~\ref{fig:eq1-timing-fidelity}c shows why the difference uses two large delays. The reference model predicts a duration that is the largest of three terms, only one of which carries the injected delay. Below a few hundred microseconds that term is not the largest, so a longer delay leaves the prediction unchanged and nothing is recovered; the difference then reflects run-to-run drift in the part of the pipeline the delay never reaches. The y axis of Figure~\ref{fig:eq1-timing-fidelity}(c) is the fraction of the requested delay that reappears in the measured duration: the mean over a cell's 36 block pairs of the delayed transfer's span minus the zero-delay transfer's span, divided by that cell's requested delay. The model predicts $-692.4$ and $-751.1$\,ns for the two cells where it is most negative, against observed $-681.8$ and $-923.6$\,ns. Five of the six negative points are not distinguishable from zero at their published 95\% intervals. The 1,000 and 200\,$\mu$s injections both sit above the work term, and their work terms are nearly equal, so the difference cancels them. Figure~\ref{fig:eq1-timing-fidelity}b gives that difference for each of the 216 measured transfers. The delivered increment has a median about 400\,ns below the request, with the middle 90\% of deliveries between $-4800$ and $+4160$\,ns. All three panels measure the injection.
\subsection{EQ2: Thermal Coupling}
\label{subsec:eval-eq2}\label{subsec:eval-thermal}
An HBF stack shares one package with the HBM region and the compute die, so the stack's own temperature does not decide how much read service it can sustain. The coupling has to be simulated transiently, and the binding component identified.

One HBF stack of 16 host channels is modeled at the 8-hi stack height Speed Grade~1 allows, with read power of 5\,W idle plus $8.4\times10^{-11}$\,J per byte, 37.256\,W at 384\,GB/s, driving a 3D-ICE package operator over the compute die, the HBM region and the eight HBF layers, stepped at 10\,ms.

Three model assumptions fix the operating point. The 384\,GB/s target is the per-cube maximum the OCP specification gives Speed Grade~1 in Table~4 (Speed Grades~2 and~3 raise it to 1.536 and 3.072\,TB/s at a 16-hi stack height, which this work does not model); the host channel is 64 bits, the value the specification fixes; and the ECC engines per channel are provisioned so that error correction does not cap the rated bandwidth, a quantity the specification leaves open.

Over horizons of 600, 3,600 and 86,400\,s, five control policies---a baseline with no throttling, the hysteresis policy at two settings, a fixed admission cap, and a proportional-integral controller---each deliver the full 384.0\,GB/s measured in the $[30,60)$\,s window, and none throttles: the seconds above the Severe and Shutdown thresholds are zero in every run, and the peak temperature is 79.8\,$^\circ$C, within $3\times10^{-5}$\,K of what the thermal fit predicts.

Within the package-safe envelope evaluated here, the package's thermal limit is reached at the compute die, not at HBF: 86.6\,W at the 90\,$^\circ$C compute limit against 121.4\,W at the HBM limit and 105.9\,W at the HBF limit, both 105\,$^\circ$C. At the compute-die limit the eight HBF layers span 44.4\,K, from 91.6\,$^\circ$C in layer 0 at the package base to 47.2\,$^\circ$C in layer 7 at the top (Figure~\ref{fig:eq2-layer-heating}).

In Figure~\ref{fig:eq2-layer-heating}, panel (a) is the first 3.0\,s at the whole-package load, and panel (b) sweeps the external heat load from 0 to 140\,W. Panels (a) and (b) are modeled predictions from a reconstructed thermal operator, not measurements of a running device. Panel (b) carries two marks: the solid line sits at 65.09\,W, the load up to which layer 0 stays flat, and the dashed line sits at the whole-package operating point, the load at which the compute die reaches its 90\,$^\circ$C ceiling in this modeled scenario. Layer 7 is still flat at 130\,W. At 0\,W in panel (b) the HBF stack is the only heat source: the compute die and the HBM region add no heat there, while the stack's own read power is present at every load.

Across seven compute phases, measured board power on the evaluation GPU (Figure~\ref{fig:eq2-physical-reference}) spans 314.988 to 431.370\,W, and the GPU's own sensor stays between about 46 and 52\,$^\circ$C. The board-power difference between the DRAM-resident and L2-resident halves of a phase rises approximately linearly with bandwidth and reaches 45.707\,W at 383.649\,GB/s. The least-squares line is drawn over all six bandwidth phases, including those that did not reach a plateau, so it describes the observations rather than a steady state taken from the plateau phases alone. No temperature here is a memory-chip temperature, since the card exposes none by any route and both NVML fields that might carry one were unavailable in all 3,877 samples. The panel (b) difference is not GDDR chip power: the two halves differ in where the buffer resides, so the difference contains whatever else changes with that residency. The 383.649\,GB/s point is GDDR7 traffic through the card's own memory system, not HBF service, and the modeled 384\,GB/s HBF target shares only that value. Panel (a)'s board-power axis and the 0 to 140\,W external heat load of Figure~\ref{fig:eq2-layer-heating}(b) are different quantities: one is a measured board draw, the other heat injected into a package model. 86.6\,W keeps its meaning as the compute-die thermal limit reverse-solved at the 90\,$^\circ$C modeled ceiling, not a card power limit; the card's own limits are 300 to 600\,W.

The service simulation runs on a 224-state model. The 224-state model and the full-order model differ by one sample in where each applies its input; with that alignment, all eleven outputs across the three input shapes agree to 0.0308\,K at worst, the \texttt{step} trace against its own 44.057\,K swing, 0.0699\% of it. At the first sample the two implementations are provably in the same state, 30.000\,$^\circ$C on all eleven channels, and the entire discrepancy at that sample equals the reduced model's feedthrough term, $D\,u[0]$, to within $2.5\times10^{-14}$\,K. Shifting the reduced model by that one sample lowers the first-sample discrepancy from 2.871, 0.7177 and 4.1406\,K to 0.0159, 0.0040 and 0.0215\,K, and 0.0215\,K is the worst of the three traces. The 3D-ICE package operator is linear, so the agreement at the comparison's own power level carries to other amplitudes of the same inputs. The comparison is model against model, not against silicon.

The three temperature limits are our own choices: 105\,$^\circ$C for HBF, the specification's operating junction range; 105\,$^\circ$C for HBM, labelled TOPER in a Micron HBM3E brief that does not say which temperature TOPER is; and 90\,$^\circ$C for the compute die, NVIDIA's published maximum for a GeForce RTX 5090 rather than the server part. The throttling triggers are configuration constants the specification leaves implementation specific. An existing approach~\cite{li2026hbfsucks} solves steady state for one stack outside the service simulator and throttles by turning off the hottest plane; the chain here is transient, whole-package, and coupled to the service side in a single run.
\subsection{EQ3: Capacity and Demand}
\label{subsec:eval-eq3}\label{subsec:eval-capacity}
Figure~\ref{fig:eq3-alloc} gives three layouts under one package-area budget of 861.1\,mm$^2$, the area the six-HBM reference layout occupies: A (six HBM stacks, no HBF), B (three HBM, two HBF) and C (one HBM, three HBF). The budget is a project assumption rather than an independently sourced package specification, and the three layouts are representative points rather than a sweep. Fitting inside the budget is a necessary condition on two-dimensional placement only: it settles nothing about routing, power delivery, PHY placement or manufacturability. A fills the budget exactly, B uses 837.7\,mm$^2$ and C 754.2\,mm$^2$; B's remaining 23.4\,mm$^2$ is a strip 1.94\,mm wide, too narrow for any device, so under a bounding-box measure B occupies 99.98\% of the budget.

512\,GiB per HBF unit is the specification's cube, while 36\,GiB per HBM stack is the specification's worked example~\cite{ocp2026hbfspec}, illustrative rather than normative, and the ranking below depends on that figure. Installed capacity rises from 216\,GiB at A to 1,572\,GiB at C, while HBM-resident key-value capacity falls from $\sim$1.72M tokens at A to $\sim$0.54M at B (Figure~\ref{fig:eq3-alloc}). Both token figures assume the complete 54.000\,GiB expert set already resident in HBM, with the remainder of HBM holding key-value state, and both are an aggregate resident mix rather than one request's context length.

At C the expert set does not fit in HBM. Of the 36\,GiB per HBM stack, 5.0121\,GiB is not available to the expert set: 2.8705\,GiB of pinned non-expert weights, 2.0000\,GiB of workspace plus safety reserve, 0.001\,GiB of metadata, and 0.1406\,GiB of transfer staging that exists only because an HBF unit is present. What remains at layout C is 30.9879\,GiB, 57.4\% of the 54.000\,GiB expert set, at zero key-value state. The workspace and staging terms are assumptions in the generating code, not measured or device figures, and the pinned weights are the only term with a measured basis. Panel (b) draws C as a short neutral hatched marker rather than a token count, and that marker carries no key-value value. What is infeasible at C is full HBM residency for the expert set, not the configuration.

The same 5.0121\,GiB is why the per-stack expert-residency threshold is 59.012\,GiB rather than 54.000\,GiB. At the other end of the same budget, MoE routing concentrates demand on a shared subset of experts, so HBF read demand per token falls with batch size.

\begin{figure}[t]
\centering
\includegraphics[width=\columnwidth]{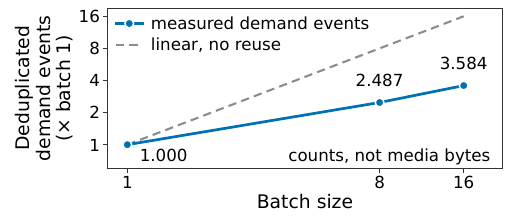}
\caption{\textbf{MoE expert demand across batch sizes.} Deduplicated demand events from a Qwen3-30B-A3B routing trace, normalized to batch 1, on logarithmic axes.}
\label{fig:eq3-demand}
\end{figure}

A captured Qwen3-30B-A3B routing trace at batches 1, 8 and 16 (48 expert layers, 128 experts, top-8) shows routing decisions growing exactly linearly, to 1, 8 and 16 times batch 1, while deduplicated demand events grow only to 2.487 and 3.584 times (Figure~\ref{fig:eq3-demand}); the dashed line in the figure repeats the routing decisions as the no-reuse reference. These are event counts rather than media bytes, and only those three batch sizes were measured. This work does not measure the timing benefit of prefetching.
\endgroup

\section{Related Work}
\label{sec:related}

\noindent\textbf{Storage and flash simulators.}
MQSim~\cite{tavakkol2018mqsim} and SimpleSSD~\cite{jung2018simplessd} model the
flash medium together with the surrounding block device and host driver stack;
both intercept at the host's block device interface.
\sys reuses the flash media model of MQSim and drives that media model in the
opposite direction: the MQSim event queue is advanced on demand while the
workload runs (Section~\ref{sec:bg}). MQSim is therefore a component of \sys
rather than a competitor: \sys charges the modeled time of a page read to one
access made by a kernel that is executing on the GPU. \sys intercepts at the
GPU's instruction path, so the requests come from a workload that is executing.
SwarmIO~\cite{kim2026swarmio} emulates a drive that GPU threads read themselves,
and meets the same conflict \chref{ch:timing} states between a timing model that
is accurate and the cost of evaluating that model at high request rates. The
drive SwarmIO emulates is reached over NVMe submission queues and a PCIe link,
so an emulator can act at a queue, a doorbell register or a replaceable driver
layer. An HBF access is issued by the GPU's own load/store units and is served
inside the accelerator package, so the host side offers no queue, no doorbell
register and no replaceable driver layer (\chref{ch:intercept}).

\noindent\textbf{GPU simulation and GPU instrumentation.}
GPGPU-Sim~\cite{bakhoda2009gpgpusim} and Accel-Sim~\cite{khairy2020accelsim}
run a modeled GPU, not the GPU on which the system will be deployed.
eGPU~\cite{yang2025egpu} places code inside a running GPU kernel and is the
direct predecessor of the instrumentation route \sys uses; \sys applies that
ability to the timing and the capacity behaviour of a device that does not exist
yet.

\noindent\textbf{Simulating a device that cannot be obtained.}
Three systems at other tiers simulate a device that cannot be obtained, and \sys
differs from each on three axes.
Cylon~\cite{yoon2026cylon} emulates a CXL-attached solid-state drive presented as
a byte-addressable memory tier, and acts on the CPU side through virtualization.
CXLMemSim~\cite{yang2026cxlmemsim} simulates a pooled memory
tier by attaching to an unmodified running program, observing memory accesses
with hardware performance counters and kernel instrumentation, and injecting
delay once per fixed interval of execution.
FlashAccel~\cite{wang2026flashaccel} is a hardware--software co-design of an
inference system built on HBF, evaluated with an event-driven simulator.
The device \sys simulates is flash inside the accelerator package, read at 64
bytes and written at 4\,KiB, with the wear and retention limits a DRAM pool does
not have. Accesses are observed and delayed in rewritten intermediate code on the
GPU, not at the virtual machine layer or in CPU performance counters. The
third axis is the loop from temperature to reliability: none of the three makes
temperature part of the device interface, whereas the HBF specification defines
temperature as values the host reads, thresholds the host sets and states the
host has to respond to.  Temperature itself is not new to simulation: prior work
extends architectural sampling to the thermal domain, because accurate
estimation of thermal behaviour is itself
time-consuming~\cite{ardestani2012thermal}.

\noindent\textbf{Full-system simulation before the silicon exists.}
A full-system simulator that already supports a new memory technology is the
obvious alternative to \sys, and the history of HBM support in
gem5~\cite{binkert2011gem5} shows what such support consists of while the parts
are still unavailable. The HBM configurations added to gem5 in September 2015
are timing values fed to the ordinary DDR controller; the commit message gives
``Timing extrapolated from existing LPDDR configurations'' as one of three
sources for those values, and adds, ``Will adjust once specs become
available.'' A memory controller written for HBM itself, with pseudo channels,
arrived only in May 2022, six years after HBM2 shipped with the P100. Before the
silicon exists, the answer a full-system simulator gives is therefore itself
extrapolated from a neighbouring standard.

\section{Discussion}
\label{sec:discuss}
The calibrated curve is measured on a complete host paging path, a software
route that presents an NVMe solid-state drive as memory through the device node
\texttt{/dev/vmem0}.  One access along that route carries a page fault, the
driver, page-cache insertion and a user-space scan, so the profile is named an
end-to-end vmem model rather than a media model.  The curve is fitted at 1, 4,
16, 64, 256 and 512 pages, and past 512 pages continues on the slope of the last
measured segment.  Random, reverse, repeated and cross-operation accesses reset
the burst length to one and are charged the single-page cost of 11,133\,ns, so
the non-linear rise measured between 64 and 512 pages shapes sequential bursts
alone.

An HBF stack inside the package pays no page fault, no driver crossing and no
page-cache insertion, and the second timing source is built for the difference:
foreground requests enter an online MQSim media model on the host along the
reference path.  A question about the medium is therefore answered with the
media model, and a question about a host-attached deployment with the calibrated
end-to-end curve.

Coverage is a property of each run rather than a fixed ceiling.
Kernels that arrive already built are covered at the level of the whole program,
where injected modeled time enters the end-to-end time and the output is
unchanged.  Results therefore carry to accesses inside registered ranges of
rewritable kernels, and widening that set is a question of registration and of
the rewriting layer, not of the timing model.

Hardware parameters that move are data rather than code.  The three synthetic
timing profiles are \texttt{conservative.json}, \texttt{nominal.json} and
\texttt{aggressive.json}; all three share a capacity of 1\,TiB and differ in
bandwidth, which ranges from 128 to 1,000\,GB/s, while the published
specification gives about 3.0\,TB/s and 512\,GB per
stack~\cite{ocp2026hbfspec}, so re-anchoring the profiles to the published
figures edits a data file.  The thermal measurements come from one GPU and one
Dell CD8P solid-state drive, since HBF parts cannot yet be obtained, and the
\texttt{simulated-warning} profile crosses the configured GPU and SSD warning
points at 30\,s and 42\,s by extrapolation, with no physical device taken there,
so figures using that profile mark the two crossings as computed values.
Retention deadlines follow the Arrhenius equation anchored on $E_a = 1.04$\,eV,
measured for 3D NAND after 10,000 program/erase cycles~\cite{luo2018heatwatch};
a medium with a different activation energy moves every deadline and the refresh
traffic each deadline forces, which is why a run reports a sweep over activation
energy rather than one lifetime.

\section{Conclusion}
\label{sec:concl}
\sys is the first open-source HBF simulator with the workload under evaluation
executing on a real GPU.  \sys rewrites the workload's PTX, routes accesses
inside a registered address range into the simulator, and supports asynchronous
TMA transfers and capacities beyond physical GPU memory.  The modeled wait
falls at the instruction that reads the returned value, not at the point of
issue.  The timing model is calibrated from measurements on real hardware
rather than parameter-sheet values.  A coupled thermal module puts HBF, HBM,
and the GPU in one package, so junction temperature changes the service rate
and the refresh traffic that the retention deadline forces.  \sys leaves the
model's run unaffected across ordinary-memory, TMA, and capacity-mode tests.
The delay \sys injects matches the delay requested to within 0.152\%.  A run
whose addresses span 110\,GiB through a 2\,GiB page cache returns a checksum
identical to a native run's.  Over horizons of 600, 3,600 and 86,400\,s, five
control policies sustain the full 384.0\,GB/s without throttling.  Within that
envelope the modeled package reaches its thermal limit at the compute die, at
86.6\,W, not at HBF.  Under one package-area budget, installed capacity rises
from 216\,GiB in an all-HBM layout to 1,572\,GiB in the most HBF-heavy one while
the key-value state HBM can hold falls.  In a captured Qwen3-30B routing trace,
deduplicated expert demand grows to 2.487 and 3.584 times batch one at batches
8 and 16, while routing decisions grow linearly.  Before HBF parts sample, \sys
lets someone designing an inference system measure a capacity or placement
decision under a real workload instead of assuming one.

\bibliographystyle{plain}
\bibliography{refs}

\begin{thebibliography}{10}

\bibitem{agrawal2026hbfaicompute}
Anurag Agrawal and Radhakrishna Giduthuri.
\newblock {HBF} in {AI} compute: A system architect's view.
\newblock Tutorial 1 (Memory technology), Hot Chips 2026, August 2026.
\newblock \url{https://hc2026.hotchips.org/}.

\bibitem{ardestani2012thermal}
Ehsan~K. Ardestani, Elnaz Ebrahimi, Gabriel Southern, and Jose Renau.
\newblock Thermal-aware sampling in architectural simulation.
\newblock In {\em Proceedings of the 2012 ACM/IEEE International Symposium on
  Low Power Electronics and Design (ISLPED)}, pages 33--38, 2012.
\newblock \url{http://masc.soe.ucsc.edu/docs/islped12.pdf}.

\bibitem{bakhoda2009gpgpusim}
Ali Bakhoda, George~L. Yuan, Wilson W.~L. Fung, Henry Wong, and Tor~M. Aamodt.
\newblock Analyzing {CUDA} workloads using a detailed {GPU} simulator.
\newblock In {\em 2009 IEEE International Symposium on Performance Analysis of
  Systems and Software (ISPASS)}, pages 163--174, 2009.

\bibitem{ocp2026hbfspec}
Chinnakrishnan Ballapuram and Dongsop Lee.
\newblock High bandwidth flash ({HBF}) high-level base die specification,
  version 0.7.0.
\newblock Open Compute Project specification, August 2026.

\bibitem{binkert2011gem5}
Nathan Binkert, Bradford Beckmann, Gabriel Black, Steven~K. Reinhardt, Ali
  Saidi, Arkaprava Basu, Joel Hestness, Derek~R. Hower, Tushar Krishna, Somayeh
  Sardashti, Rathijit Sen, Korey Sewell, Muhammad Shoaib, Nilay Vaish, Mark~D.
  Hill, and David~A. Wood.
\newblock The {gem5} simulator.
\newblock {\em ACM SIGARCH Computer Architecture News}, 39(2):1--7, 2011.
\newblock Source repository: \url{https://github.com/gem5/gem5}.

\bibitem{grattafiori2024llama3}
Aaron Grattafiori et~al.
\newblock The {Llama} 3 herd of models.
\newblock arXiv:2407.21783, 2024.
\newblock \url{https://arxiv.org/abs/2407.21783}.

\bibitem{ha2026h3}
Minho Ha, Euiseok Kim, and Hoshik Kim.
\newblock {H\textsuperscript{3}}: Hybrid architecture using high bandwidth
  memory and high bandwidth flash for cost-efficient {LLM} inference.
\newblock {\em IEEE Computer Architecture Letters}, 25(1):49--52, 2026.

\bibitem{huerta2025parallel}
Rodrigo Huerta and Antonio Gonz{\'a}lez.
\newblock Parallelizing a modern {GPU} simulator.
\newblock In {\em 2nd Workshop on Computer Architecture Modeling and Simulation
  (CAMS), co-located with MICRO}, 2024.
\newblock \url{https://arxiv.org/abs/2502.14691}.

\bibitem{jedec2016jesd218b}
{JEDEC Solid State Technology Association}.
\newblock Solid-state drive ({SSD}) requirements and endurance test method.
\newblock JEDEC Standard JESD218B, revision of JESD218A of February 2011, March
  2016.

\bibitem{jedec2025hbm4}
{JEDEC Solid State Technology Association}.
\newblock {JEDEC} and industry leaders collaborate to release {JESD270-4}
  {HBM4} standard: Advancing bandwidth, efficiency, and capacity for {AI} and
  {HPC}, April 2025.
\newblock
  \url{https://www.jedec.org/news/pressreleases/jedec\%C2\%AE-and-industry-leaders-collaborate-release-jesd270-4-hbm4-standard-advancing}.

\bibitem{jouppi2021tpuv4i}
Norman~P. Jouppi, Doe~Hyun Yoon, Matthew Ashcraft, Mark Gottscho, Thomas~B.
  Jablin, George Kurian, James Laudon, Sheng Li, Peter Ma, Xiaoyu Ma, Thomas
  Norrie, Nishant Patil, Sushma Prasad, Cliff Young, Zongwei Zhou, and David
  Patterson.
\newblock Ten lessons from three generations shaped {Google}'s {TPUv4i}:
  Industrial product.
\newblock In {\em Proceedings of the 48th Annual International Symposium on
  Computer Architecture (ISCA)}, pages 1--14, 2021.

\bibitem{jouppi2017tpu}
Norman~P. Jouppi, Cliff Young, Nishant Patil, David Patterson, et~al.
\newblock In-datacenter performance analysis of a tensor processing unit.
\newblock In {\em Proceedings of the 44th Annual International Symposium on
  Computer Architecture (ISCA)}, pages 1--12, 2017.

\bibitem{ju2026tilelens}
Jae~Hyung Ju, Euijun Chung, Hritvik Taneja, Anish Saxena, Shinnung Jeong,
  Hyesoon Kim, and Moinuddin~K. Qureshi.
\newblock {TileLens}: Efficiently using large-granularity memory systems with
  transparent two-dimensional memory layout.
\newblock arXiv:2607.04031, 2026.
\newblock \url{https://arxiv.org/abs/2607.04031}.

\bibitem{jung2025hbfwinner}
Ilju Jung.
\newblock {KAIST} professor joungho kim: ``an era is coming in which {HBF}
  decides the memory winner''.
\newblock THE ELEC, September 2025.
\newblock \url{https://www.thelec.kr/news/articleView.html?idxno=40242}.

\bibitem{jung2018simplessd}
Myoungsoo Jung, Jie Zhang, Ahmed Abulila, Miryeong Kwon, Narges Shahidi, John
  Shalf, Nam~Sung Kim, and Mahmut Kandemir.
\newblock {SimpleSSD}: Modeling solid state drives for holistic system
  simulation.
\newblock {\em IEEE Computer Architecture Letters}, 17(1):37--41, 2018.
\newblock arXiv:1705.06419, \url{https://arxiv.org/abs/1705.06419}.

\bibitem{juravsky2024hydragen}
Jordan Juravsky, Bradley Brown, Ryan Ehrlich, Daniel~Y. Fu, Christopher R{\'e},
  and Azalia Mirhoseini.
\newblock {Hydragen}: High-throughput {LLM} inference with shared prefixes.
\newblock arXiv:2402.05099, 2024.
\newblock \url{https://arxiv.org/abs/2402.05099}.

\bibitem{kennedy2026oxmiqhbf}
Patrick Kennedy.
\newblock Oxmiq labs {HBF} in {AI} compute at hot chips 2026.
\newblock ServeTheHome, August 2026.
\newblock
  \url{https://www.servethehome.com/oxmiq-labs-hbf-in-ai-compute-at-hot-chips-2026/}.

\bibitem{khairy2020accelsim}
Mahmoud Khairy, Zhesheng Shen, Tor~M. Aamodt, and Timothy~G. Rogers.
\newblock {Accel-Sim}: An extensible simulation framework for validated {GPU}
  modeling.
\newblock In {\em Proceedings of the 47th Annual International Symposium on
  Computer Architecture (ISCA)}, 2020.
\newblock
  \url{https://people.ece.ubc.ca/aamodt/publications/papers/accelsim.isca2020.pdf}.

\bibitem{kim2026swarmio}
Hyeseong Kim, Gwangoo Yeo, and Minsoo Rhu.
\newblock {SwarmIO}: Towards 100 million {IOPS} {SSD} emulation for
  next-generation {GPU}-centric storage systems.
\newblock arXiv:2604.06668, 2026.
\newblock \url{https://arxiv.org/abs/2604.06668}.

\bibitem{kouchi2021flash}
Toshiyuki Kouchi, Mami Kakoi, Noriyasu Kumazaki, Akio Sugahara, Akihiro
  Imamoto, Yasufumi Kajiyama, Yuri Terada, Bushnaq Sanad, Naoaki Kanagawa,
  Takuyo Kodama, Ryo Fukuda, Hiromitsu Komai, Norichika Asaoka, Hidekazu
  Ohnishi, Ryosuke Isomura, Takaya Handa, Kensuke Yamamoto, Yuki Ishizaki, Yoko
  Deguchi, Atsushi Okuyama, Junichi Sato, Hiroki Yabe, Hua-Ling~Cynthia Hsu,
  and Masahiro Yoshihara.
\newblock A 128{Gb} 1-bit/cell 96-word-line-layer {3D} flash memory to improve
  the random read latency with {$t_{\mathrm{Prog}}$} = 75\,$\mu$s and
  {$t_{\mathrm{R}}$} = 4\,$\mu$s.
\newblock {\em IEEE Journal of Solid-State Circuits}, 56(1):225--234, 2021.

\bibitem{kwon2026hbf2038}
Dongjun Kwon.
\newblock ``father of {HBM}'' professor joungho kim: ``the {HBF} market will
  exceed {HBM} in 2038''.
\newblock ETNews, February 2026.
\newblock \url{https://www.etnews.com/20260203000211}.

\bibitem{li2018femu}
Huaicheng Li, Mingzhe Hao, Michael~Hao Tong, Swaminathan Sundararaman, Matias
  Bj{\o}rling, and Haryadi~S. Gunawi.
\newblock The {CASE} of {FEMU}: Cheap, accurate, scalable and extensible flash
  emulator.
\newblock In {\em 16th USENIX Conference on File and Storage Technologies
  (FAST)}, 2018.
\newblock \url{https://www.usenix.org/conference/fast18/presentation/li}.

\bibitem{li2026hbfsucks}
Zhuoran Li, Zhuohang Bian, Xin Huang, Yibo Zhao, Guangyu Sun, and Youwei Zhuo.
\newblock {HBF} sucks? a full-stack characterization of high-bandwidth flash
  for {KV}-centric {LLM} serving.
\newblock arXiv:2608.11668, 2026.
\newblock \url{https://arxiv.org/abs/2608.11668}.

\bibitem{liao2019davinci}
Heng Liao, Jiajin Tu, Jing Xia, and Xiping Zhou.
\newblock {DaVinci}: A scalable architecture for neural network computing.
\newblock Hot Chips 31 Symposium on High-Performance Chips, August 2019.

\bibitem{luo2024hopper}
Weile Luo, Ruibo Fan, Zeyu Li, Dayou Du, Qiang Wang, and Xiaowen Chu.
\newblock Benchmarking and dissecting the {Nvidia} {Hopper} {GPU} architecture.
\newblock In {\em 2024 IEEE International Parallel and Distributed Processing
  Symposium (IPDPS)}, pages 656--667, 2024.
\newblock arXiv:2402.13499, \url{https://arxiv.org/abs/2402.13499}.

\bibitem{luo2018heatwatch}
Yixin Luo, Saugata Ghose, Yu~Cai, Erich~F. Haratsch, and Onur Mutlu.
\newblock {HeatWatch}: Improving {3D} {NAND} flash memory device reliability by
  exploiting self-recovery and temperature awareness.
\newblock In {\em 2018 IEEE International Symposium on High Performance
  Computer Architecture (HPCA)}, pages 504--517, 2018.
\newblock
  \url{https://www.cs.cmu.edu/~yixinluo/index_files/heatwatch_hpca18.pdf}.

\bibitem{micron2026form10q}
{Micron Technology, Inc.}
\newblock Quarterly report on {Form 10-Q} for the quarterly period ended may
  28, 2026.
\newblock U.S. Securities and Exchange Commission, 2026.
\newblock Accession 0000723125-26-000015, filed June 25, 2026;
  \url{https://www.sec.gov/Archives/edgar/data/723125/000072312526000015/mu-20260528.htm}.

\bibitem{nvidia2024h100}
{NVIDIA Corporation}.
\newblock {NVIDIA} {H100} {Tensor} {Core} {GPU} datasheet.
\newblock Product datasheet, document 3440270, September 2024.
\newblock
  \url{https://resources.nvidia.com/en-us-gpu-resources/h100-datasheet-24306}.

\bibitem{paek2026fatherhbm}
Jongmin Paek.
\newblock ``father of {HBM}'' aims to secure {HBF} patents.
\newblock Asia Economy, English edition, February 2026.
\newblock \url{https://view.asiae.co.kr/en/article/2026021009235328565}.

\bibitem{pan2026nextgen}
Junrui Pan, Weili An, Cesar~Avalos Baddouh, Christin~David Bose, Ni~Kang, Aaron
  Barnes, Ahmad Alawneh, Fangjia Shen, Yechen Liu, Anusuya Nallathambi, Atthin
  Chandrashekar, and Timothy~G. Rogers.
\newblock Architecting the next generation of asynchronous, distributed {GPU}s
  for the {AI} era.
\newblock arXiv:2608.22602, 2026.
\newblock \url{https://arxiv.org/abs/2608.22602}.

\bibitem{petrucci2026hbfallyouneed}
Vinicius Petrucci, Felippe Zacarias, and Vishal Tanna.
\newblock Is high-bandwidth flash all you need?
\newblock In {\em HotInfra '26, co-located with ISCA '26}, 2026.
\newblock \url{https://hotinfra.org/2026/papers/hotinfra26-final83.pdf}.

\bibitem{pope2022scaling}
Reiner Pope, Sholto Douglas, Aakanksha Chowdhery, Jacob Devlin, James Bradbury,
  Anselm Levskaya, Jonathan Heek, Kefan Xiao, Shivani Agrawal, and Jeff Dean.
\newblock Efficiently scaling transformer inference.
\newblock In {\em Proceedings of Machine Learning and Systems}, volume~5, pages
  606--624, 2023.
\newblock arXiv:2211.05102, \url{https://arxiv.org/abs/2211.05102}.

\bibitem{sandisk2025factsheet}
{Sandisk Corporation}.
\newblock {HBF} fact sheet: Sandisk unveils the future of memory architecture
  for {AI}.
\newblock Tech brief, July 2025.
\newblock
  \url{https://documents.sandisk.com/content/dam/asset-library/en_us/assets/public/sandisk/collateral/company/Sandisk-HBF-Fact-Sheet.pdf}.

\bibitem{sandisk2026hbfspecrelease}
{Sandisk Corporation} and {SK hynix}.
\newblock Sandisk and {SK hynix} advance global standardization of high
  bandwidth flash with release of first {OCP} technical specification, August
  2026.
\newblock
  \url{https://www.sandisk.com/company/newsroom/press-releases/2026/2026-08-03-Sandisk-and-sk-hynix-advance-global-standardization-of-hbf}.

\bibitem{sano2023cxlgpu}
Shintaro Sano, Yosuke Bando, Kazuhiro Hiwada, Hirotsugu Kajihara, Tomoya
  Suzuki, Yu~Nakanishi, Daisuke Taki, Akiyuki Kaneko, and Tatsuo Shiozawa.
\newblock {GPU} graph processing on {CXL}-based microsecond-latency external
  memory.
\newblock In {\em Proceedings of the SC '23 Workshops of the International
  Conference on High Performance Computing, Network, Storage, and Analysis
  (SC-W)}, 2023.
\newblock \url{https://arxiv.org/abs/2312.03113}.

\bibitem{sheng2023flexgen}
Ying Sheng, Lianmin Zheng, Binhang Yuan, Zhuohan Li, Max Ryabinin, Beidi Chen,
  Percy Liang, Christopher R{\'e}, Ion Stoica, and Ce~Zhang.
\newblock {FlexGen}: High-throughput generative inference of large language
  models with a single {GPU}.
\newblock In {\em Proceedings of the 40th International Conference on Machine
  Learning (ICML)}, volume 202 of {\em Proceedings of Machine Learning
  Research}, pages 31094--31116, 2023.
\newblock \url{https://proceedings.mlr.press/v202/sheng23a.html}.

\bibitem{sun2019mgpusim}
Yifan Sun, Trinayan Baruah, Saiful~A. Mojumder, Shi Dong, Xiang Gong, Shane
  Treadway, Yuhui Bao, Spencer Hance, Carter McCardwell, Vincent Zhao, Harrison
  Barclay, Amir~Kavyan Ziabari, Zhongliang Chen, Rafael Ubal, Jos{\'e}~L.
  Abell{\'a}n, John Kim, Ajay Joshi, and David Kaeli.
\newblock {MGPUSim}: Enabling multi-{GPU} performance modeling and
  optimization.
\newblock In {\em Proceedings of the 46th International Symposium on Computer
  Architecture (ISCA)}, pages 197--209, 2019.

\bibitem{tavakkol2018mqsim}
Arash Tavakkol, Juan G{\'o}mez-Luna, Mohammad Sadrosadati, Saugata Ghose, and
  Onur Mutlu.
\newblock {MQSim}: A framework for enabling realistic studies of modern
  multi-queue {SSD} devices.
\newblock In {\em 16th USENIX Conference on File and Storage Technologies
  (FAST)}, 2018.
\newblock \url{https://www.usenix.org/conference/fast18/presentation/tavakkol}.

\bibitem{villa2019nvbit}
Oreste Villa, Mark Stephenson, David Nellans, and Stephen~W. Keckler.
\newblock {NVBit}: A dynamic binary instrumentation framework for {NVIDIA}
  {GPUs}.
\newblock In {\em Proceedings of the 52nd Annual IEEE/ACM International
  Symposium on Microarchitecture (MICRO)}, pages 372--383, 2019.
\newblock
  \url{https://d1qx31qr3h6wln.cloudfront.net/publications/MICRO_2019_NVBit.pdf}.

\bibitem{wahlgren2025mi300a}
Jacob Wahlgren, Gabin Schieffer, Ruimin Shi, Edgar~A. Le{\'o}n, Roger Pearce,
  Maya Gokhale, and Ivy Peng.
\newblock Dissecting {CPU-GPU} unified physical memory on {AMD} {MI300A}
  {APUs}.
\newblock In {\em 2025 IEEE International Symposium on Workload
  Characterization (IISWC)}, 2025.
\newblock arXiv:2508.12743, \url{https://arxiv.org/abs/2508.12743}.

\bibitem{wang2026flashaccel}
Xinyu Wang, Yalong Xue, Xiaotian Sun, Xiaoyu Zhang, Xinjiang Zhang, Chunmeng
  Dou, Xueqi Li, and Xiaoming Chen.
\newblock {FlashAccel}: Leveraging high-bandwidth flash ({HBF}) for
  high-throughput {LLM} inference.
\newblock arXiv:2607.10186, 2026.
\newblock \url{https://arxiv.org/abs/2607.10186}.

\bibitem{yang2026cxlmemsim}
Yiwei Yang, Shri~Vishakh Devanand, Brian Zhao, Yusheng Zheng, Pooneh
  Safayenikoo, Tanvir~Ahmed Khan, and Andi Quinn.
\newblock {CXLMemSim}: Practical performance simulation and characterization of
  {CXL} 3.0 memory systems.
\newblock In {\em Proceedings of the 35th International Symposium on
  High-Performance Parallel and Distributed Computing (HPDC)}, pages 567--569,
  2026.

\bibitem{yang2025egpu}
Yiwei Yang, Tong Yu, Yusheng Zheng, and Andrew Quinn.
\newblock {eGPU}: Extending {eBPF} programmability and observability to {GPUs}.
\newblock In {\em 4th Workshop on Heterogeneous Composable and Disaggregated
  Systems (HCDS)}, 2025.
\newblock \url{https://asplos.dev/pdf/bpftime_super.pdf}.

\bibitem{yin2026hbfserving}
Yihan Yin, Yinlun Zhao, Zhixin Yun, Guanying Wu, Feng Zhu, Kai Tao, Shu Li, Fei
  Huang, Zhe Zhang, Shuangchen Li, and Hongzhong Zheng.
\newblock Potential applications of {HBF} in {LLM} serving systems, August
  2026.
\newblock arXiv:2608.13127, \url{https://arxiv.org/abs/2608.13127}.

\bibitem{yoon2026cylon}
Dongha Yoon, Hansen Idden, Jinshu Liu, Berkay Inceisci, Sam~H. Noh, and
  Huaicheng Li.
\newblock {Cylon}: Fast and accurate full-system emulation of {CXL-SSDs}.
\newblock In {\em 24th USENIX Conference on File and Storage Technologies
  (FAST)}, 2026.
\newblock \url{https://www.usenix.org/conference/fast26/presentation/yoon}.

\bibitem{yuan2024roofline}
Zhihang Yuan, Yuzhang Shang, Yang Zhou, Zhen Dong, Zhe Zhou, Chenhao Xue,
  Bingzhe Wu, Zhikai Li, Qingyi Gu, Yong~Jae Lee, Yan Yan, Beidi Chen, Guangyu
  Sun, and Kurt Keutzer.
\newblock {LLM} inference unveiled: Survey and roofline model insights.
\newblock arXiv:2402.16363, 2024.
\newblock \url{https://arxiv.org/abs/2402.16363}.

\bibitem{zhu2026threeice}
Kai Zhu, Darong Huang, Luis Costero, and David Atienza.
\newblock {3D-ICE} 4.0: Accurate and efficient thermal modeling for {2.5D/3D}
  heterogeneous chiplet systems.
\newblock In {\em Proceedings of the 2026 Design, Automation \& Test in Europe
  Conference (DATE)}, pages 1--7, 2026.

\end{thebibliography}

\balance

\end{document}